\documentclass[11pt]{article}
\usepackage{amsmath,amssymb,latexsym,float,epsfig,subfig}
\usepackage{latexsym}
\usepackage{natbib}
\usepackage{amsthm}
\usepackage{cases}
\usepackage{epsfig}
\usepackage{algpseudocode}
\usepackage{scalefnt}
\usepackage{booktabs}
\usepackage[english]{babel}
\usepackage{graphicx}
\usepackage{titlesec}
\usepackage{colortbl}

\usepackage{soul}
\usepackage{natbib}

\usepackage{tikz-qtree}
\usetikzlibrary{trees}

\usepackage[normalem]{ulem}
\usepackage{xargs}                      

\usepackage{float}
\usepackage{bbm}

\usepackage{epsfig}
\usepackage{enumerate}
\usepackage{caption}
\usepackage{subfig}
\usepackage{booktabs}
\usepackage[english]{babel}
\usepackage{graphicx}
\usepackage{multirow}
\usepackage{rotating}
\usepackage{fancyhdr}
\usepackage{setspace}
\usepackage[section]{placeins}
\usepackage{breqn}
\usepackage{lipsum}
\usepackage{mwe}
\usepackage{pseudocode}
\usepackage{algorithmicx}
\usepackage{algorithm}
\usepackage{url} 
\usepackage{appendix}
\usepackage{xcolor}
\usepackage{optidef}
\usepackage{graphicx}
\usepackage{caption}
\usepackage{mathtools}

\usepackage{makecell}

\topmargin by -\headsep \textheight 8.9in \oddsidemargin 0pt
\evensidemargin \oddsidemargin \marginparwidth 0.5in \textwidth
6.5in
\date{}
\usepackage{graphicx}
\usepackage{bbm}
\usepackage{mathtools}
\usepackage{graphicx}
\usepackage{caption}
\usepackage{longtable}
\usepackage{pifont}
\usepackage{hyperref}
\usepackage{tikz}
\usetikzlibrary{shapes.geometric, arrows}
\allowdisplaybreaks

\usepackage{array}
\newcolumntype{P}[1]{>{\centering\arraybackslash}p{#1}}
\begin{document}

\title{A Review on Response Strategies in Infrastructure Network Restoration}

\author{
Subhojit Biswas \thanks{subhojit.biswas@tamu.edu} \and
Bahar \c{C}avdar \thanks{cavdab2@rpi.edu} \and Joseph Geunes \thanks{geunes@tamu.edu}}
\date{%
     $^*$$^\ddagger$\small{Department of Industrial and Systems Engineering, Texas A\&M University, College Station, TX 77843}\\[2ex]%
    $^\dagger$\small{Department of Industrial and Systems Engineering, Rensselaer Polytechnic Institute, Troy, NY, 12180}%
}

\maketitle

\begin{abstract}
This paper reviews the literature on response strategies for restoring infrastructure networks in the aftermath of a disaster. Our motivation for this review is twofold. First, the frequency and magnitude of natural and man-made disasters (e.g., wild fires, tornadoes, global pandemics, terrorist attacks) have been increasing. These events disrupt the operation of infrastructure networks, preventing the delivery of vital goods and services such as power and food. Therefore, it is critical to understand the state-of-the-art in responding to network disruptions in order to develop efficient strategies to mitigate their impacts. Second, it is critical to enable timely decisions in a rapidly changing and unpredictable environment while accounting for numerous interrelated factors. Because the vast majority of response decision problems are computationally challenging, quickly finding solutions that are compatible with real-time decision making is a difficult task. Hence, it is important to understand the nature of response activities and decisions, as well as the available solution methodologies and inherent trade-offs between computation time and solution quality. We review quantitative response methodologies developed for infrastructure network restoration, classifying relevant studies based on the properties of the underlying network. In particular, we focus on resource allocation, scheduling, routing and repair efforts within the domain of power, road, and water, oil and gas network restoration.  
We also discuss open research questions and future research directions. 

\end{abstract}

\section{Introduction}
Over the past few decades, disasters worldwide have resulted in significant economic, social, and physical losses, causing considerable negative consequences for populations living in affected areas due to disruptions in the operation of critical infrastructure networks. Typical infrastructure networks, namely power, water, natural gas, telecommunications and physical road networks, are inter-connected, which allows them to function collaboratively to ensure a steady flow of necessary goods and services. Society's high level of susceptibility to various hazards is underscored by recent events including Hurricane Harvey (2017), floods in South Carolina (2015), Hurricane Sandy (2012), earthquakes in Chile (2010), New Zealand (2010-2011), and Japan (2011), the latter of which was followed by a Tsunami, and winter storms in China (2008) \citep{perrier2013survey}. In February 2021, winter storm Uri struck Texas, affecting 4.5 million homes without power for several days \citep{pollock2021gov,busby2021cascading}. 
The resulting power outages also lead to disruptions in medical services and water supply \citep{mcnamara2021over}. AccuWeather estimated the economic costs during the winter to be \$130 billion in Texas and \$150 billion nationwide as a result of production losses and physical damage to infrastructure, property, and equipment \citep{puelo2021damages}.  
These negative impacts highlight a pressing need to improve and promote resilient infrastructure by government organizations. The United Nations disaster risk reduction office recently published a report focusing on mitigating risks and enhancing resilience for critical infrastructure (such as energy, water, transport, and telecommunication) in Central Asia and Europe to address the adverse impacts of climate change and the risk associated with natural disasters \citep{united1901global}. The U.S. government has similarly placed increased emphasis on resilience planning for these critical infrastructure networks \citep{house2013critical}. 

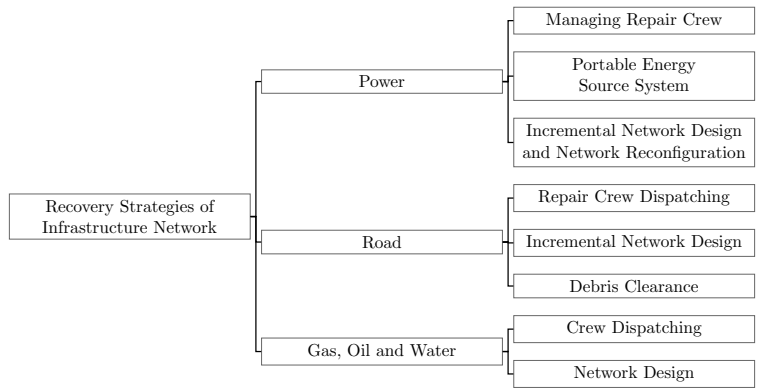
\begin{figure}[H]
\centering
\begin{tikzpicture}[level distance=2.2in,sibling distance=.15in,scale=0.6]
\tikzset{edge from parent/.style= 
            {thick, draw,
                edge from parent fork right},every tree node/.style={draw,minimum width=1.75in,text width=2in, align=center, draw=black!60, fill=white!5},grow'=right}

\Tree 
    [. {Recovery Strategies of Infrastructure Network}
        [.{Power}
                [.{Managing Repair Crew} ]
            [.{Portable Energy Source System } ]
            [.{Incremental Network Design and Network Reconfiguration } ]
                    ]
        [.{Road}
                [.{Repair Crew Dispatching } ]
            [.{Incremental Network Design} ]
            [.{Debris Clearance} ]
        ] 
        [.{Gas, Oil and Water} 
        [.{Crew Dispatching } ]
        [.{Network Design} ]
        ]
    ]
\end{tikzpicture}
\caption{Broad classification of recovery strategies for different infrastructure networks.}
    \label{flowchart}
\end{figure}

\begin{figure}[H]
\centering
    \centering
    \includegraphics[width=.75\linewidth, height =.53\linewidth]{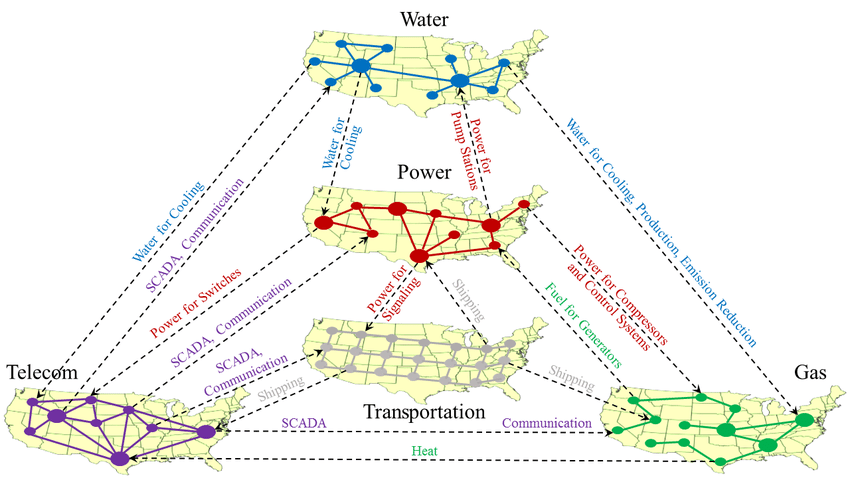}
    \caption{Interdependency between critical infrastructure networks (Source: \citealp{zuev2018reliability}).}
    \label{interdependent}
\end{figure}

Resilience planning is generally divided into two phases, a preparedness phase and a response phase. The preparedness phase takes place before the occurrence of a disaster. Preparedness actions focus on pre-disaster strategies such as positioning resources in anticipation of disruptions, capacity planning and strengthening the underlying network. The response phase takes place after a disaster and mainly focuses on the allocation of available resources (e.g., response crews, materials) to affected locations in order to restore the system with minimum negative impact. While it is critical to have efficient strategies for both phases to achieve higher levels of resilience, this review will focus on the response phase for infrastructure network resilience. Our motivation for this is twofold. First, preparedness actions generally require large investments and are spread over a long time frame, with a focus on enabling effective response in the future. Since disruptions occur without warning and even before planning-phase activities can be fully implemented, having current and efficient response strategies in place becomes critical. Second, response decisions are computationally highly demanding. A key metric that determines the performance of a response strategy is how long it takes to recover after failures. This recovery time depends on multiple factors, including how long it takes to determine a response decision and the efficiency of the resulting decision. The former is related to the computation time and the latter is a function of the quality of the solution obtained. Therefore, there is a significant trade-off between the time allocated to the computation of potential solutions and the resulting solution quality, especially in light of the real-time need for fast response actions. With this in mind, we review the literature on recovery strategies for infrastructure networks focusing on power, road, and water, gas, and oil networks using the classification shown in in Figure \ref{flowchart}. These systems of networks are highly interconnected, such that the disruption of one of them might lead to disruption of others (please see Figure \ref{interdependent}). 

This paper is organized as follows: Section \ref{Sec:RevMet} discusses the methodology used in conducting the literature review. In Section \ref{Sec:ResOv}, we summarize definitions and characterizations of resilience. In Section \ref{Sec:Power}, we review the literature on power network restoration methodologies, focusing on repair crew scheduling and routing, portable energy systems, and incremental network design and network reconfiguration. Section \ref{Sec:Road} reviews the literature on road network restoration, focusing on repair scheduling, routing, and debris clearance. In Section \ref{Sec:Others}, we summarize the combined literature on other infrastructure networks, namely water, gas and oil, as studies specifically addressing these networks are relatively sparse compared to power networks and road networks. Finally, Section \ref{Sec:Conc} discusses gaps in the literature and presents future research directions.

\section{Review Methodology}
\label{Sec:RevMet}
We carry out a systematic review of the literature on recovery techniques for critical infrastructure networks. Our review is organized around (i) power networks, (ii) road networks, and (iii) water, gas and oil networks. 
We review scholarly publications addressing the response and recovery phases of critical infrastructure network resilience. We examine papers that have undergone peer review within the last 20 years. To compile relevant literature, we conduct searches on the Google Scholar database and access publications from reputable publishers, including INFORMS, Elsevier, Springer, Taylor and Francis, and IEEE, as well as digital services such as Wiley Online Services and JSTOR. Our search strategy employs a combination of diverse keywords including ``infrastructure network resilience," ``scheduling of infrastructure networks," ``routing of infrastructure networks," ``repair of power networks," ``repair of road networks and debris clearance," ``resilience in distribution networks," ``supply resilience strategy," ``resilience," ``infrastructure network vulnerability," ``infrastructure network disruptions," ``power network disruptions," ``repair of network disruptions," and ``road network disruptions." These keywords are used during searches of various databases and the home websites of the aforementioned publishers. Upon identifying a specific paper, our evaluation process thoroughly examines its abstract, introduction, and conclusion. If the content aligns with our target area, we proceed to read the full paper; otherwise, we exclude it. For papers deemed relevant, we scrutinize their citations  to ensure a comprehensive review, in order to make every effort to ensure that all relevant literature has been exhaustively searched and reviewed. These steps provide an understanding of the existing literature within the realm of infrastructure network recovery. We have reviewed more than 150 papers, including the relevant citations within the papers we have cited. We illustrate the entire review process using the flowchart in Figure \ref{fig:Review Methodology}.

\tikzstyle{startstop} = [rectangle, rounded corners, 
minimum width=1.5cm, 
minimum height=1cm,
text centered, 
draw=black, 
fill=white!30]

\tikzstyle{io} = [trapezium, 
trapezium stretches=true, 
trapezium left angle=70, 
trapezium right angle=110, 
minimum width=3cm, 
minimum height=1cm, text centered, 
draw=black, fill=white!30]

\tikzstyle{process} = [rectangle, 
minimum width=2cm, 
minimum height=1cm, 
text centered, 
text width=3cm, 
draw=black, 
fill=white!30]

\tikzstyle{decision} = [diamond, 
text centered, 
draw=black, 
fill=white!30]
\tikzstyle{arrow} = [thick,->,>=stealth]
\begin{figure}[H]
\centering

\begin{tikzpicture}[node distance=2cm, every node/.style={scale=0.7}]

\node (start) [startstop] {Start};
\node (pro1) [process, above of=start, yshift=1.0cm] {Search keywords};
\node (dec1) [process, right of=pro1, xshift=2cm] {Examine abstracts, introductions and conclusions};
\node (pro2a) [decision, right of=dec1, xshift=1.5cm] {Relevant?};
\node (dec2) [process, right of=pro2a, xshift=1.75cm] {Read full papers };
\node (pro2c) [process, right of=dec2, xshift=2cm] { Scrutinize citations for \\ additional \\relevant \\ publications  };
\node (pro2b) [decision, right of=pro2c, xshift=2cm, align = left] {Papers \\ exhausted?};
\node (stop) [startstop, below of=pro2b, yshift= -1.0cm] {Stop};

\draw [arrow] (start) -- (pro1);
\draw [arrow] (pro1) -- (dec1);
\draw [arrow] (dec1) -- (pro2a);
\draw [arrow] (pro2a.north)node[anchor=south] {No}  -|  (dec1.north);
\draw [arrow] (pro2a) -- node[anchor=south] {Yes}(dec2);
\draw [arrow] (dec2) -- (pro2c);
\draw [arrow] (pro2c) -- (pro2b);
\draw [arrow] (pro2b.north)node[anchor=south] {No}  -| (dec1.north);
\draw [arrow] (pro2b) -- node[anchor=east]  {Yes}(stop);

\end{tikzpicture}
\caption{Flow chart of the literature review approach.}
    \label{fig:Review Methodology}
\end{figure}
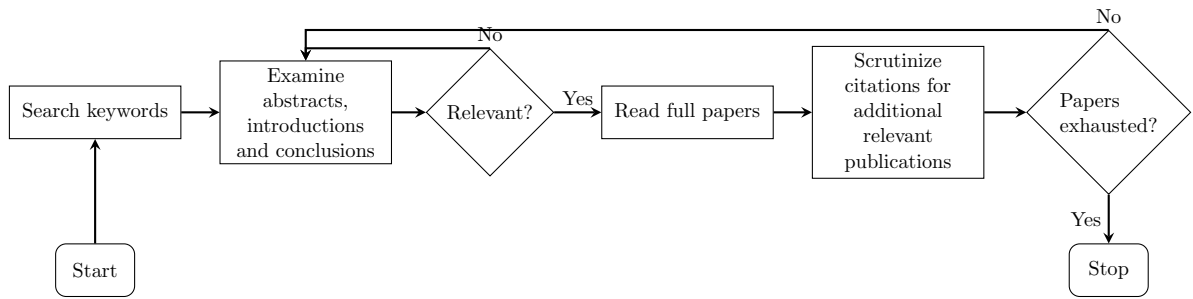

\section{Resilience Overview}
\label{Sec:ResOv}
Understanding what constitutes a resilient system and how to achieve resilience has attracted significant attention within several application areas, such as supply chains, transportation and energy, leading to a growing body of literature. Many studies have provided a definition of resilience \citep[e.g.,][]{henry2012generic,francis2014metric,ouyang2014multi,panteli2017metrics}. While these definitions vary, they have some common components. \cite{ribeiro2018supply} provide a new definition capturing these common components. According to them, a resilient system is ``one that is able to prepare, respond and recover from the disturbances and afterwards maintain a positive steady state operation in an acceptable cost and time." Based on this, higher levels of resilience require careful attention and planning before (i.e., preparation) and after (i.e., response) a disruption occurs to maintain high levels of system flexibility. 

In addition, to summarize the literature and identify gaps, many researchers provided reviews related to the resilience of systems. Among these, \cite{snyder2016or}, \cite{ribeiro2018supply} and \cite{hosseini2019review} provide comprehensive reviews detailing various quantitative approaches and techniques used to enhance and measure resilience across the entire supply chain. \cite{tukamuhabwa2015supply} examine the research methods and key strategies in supply chain management and propose systems that are intricate and adaptive in nature to study supply chain resilience. \cite{ponomarov2009understanding} review resilience using an interdisciplinary approach; they consider resilience from economic, ecological, and physiological perspectives. These reviews are characterized in Table \ref{SC Table} based on the scope and the time frame covered. In this paper, we review the literature on infrastructure network resilience, focusing on the response phase with two goals in particular. First, we emphasize the importance of real-time response decisions and the underlying decision problems, which are often combinatorial in nature and include uncertainty.
It is thus imperative to comprehend the latest advancements in making effective decisions under uncertainty within limited computational time. Second, to the best of the authors' knowledge, a comprehensive review of the disruption response phase for general infrastructure networks does not exist at this time. For studies focusing on the preparation phase for infrastructure network resilience, we refer interested readers to \cite{baroud2014importance}, \cite{d2015mathematical}, and \cite{nan2017quantitative} for independent networks, and \cite{sharkey2015interdependent}, \cite{gonzalez2016interdependent}, \cite{barker2017defining}, \cite{gonzalez2017efficient}, and  \cite{smith2020interdependent} for interdependent and interconnected networks.

In the remainder of this paper, we review response strategies for infrastructure networks within three categories, focusing on (i) power networks, (ii) road networks, and (iii) water, gas and oil networks.

\begin{table}[H]
\scalefont{0.8}
\centering
\caption{Classification of supply chain resilience (SCR) review papers. }
\label{SC Table}
\begin{tabular}{l|c | c}
\hline 
\hline
\textbf{Review paper} &\textbf{Scope}   & \textbf{Time-span }\\
\hline 
\cite{hosseini2019review} & Quantitative review in SCR   & 2000-2018 \\
\hline
\cite{ribeiro2018supply} & Quantitative review in SCR   & 2006-2017 \\
\hline
\cite{snyder2016or} & \makecell{Quantitative and qualitative reviews on SC disruptions}   & 1991-2012 \\
\hline
\cite{tukamuhabwa2015supply} & \makecell{Strategies and research methods in SC management}   & 2001-2015 \\
\hline
\cite{ponomarov2009understanding} & SCR through interdisciplinary approach  & 1991-2008 \\ \hline
\textbf{This review} & Quantitative review on infrastructure network restoration & 1998-2022 \\ 
\hline \hline
\end{tabular}
\end{table}

\normalsize

\section{Power Network Restoration}
\label{Sec:Power}

\begin{figure}[H]
\centering
\subfloat[Number of yearly electric disturbances in the U.S. between 2002-2020 \cite{econ}.]{\label{electrical dist in 2020}{\includegraphics[width=0.45\textwidth]{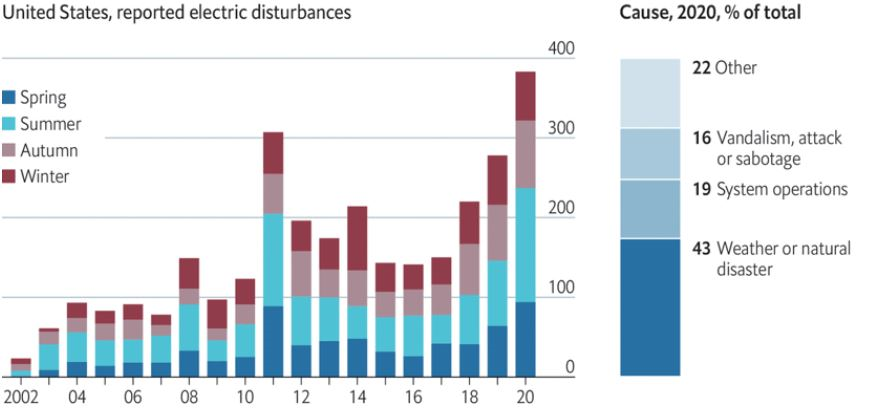}}}\hfill
\subfloat[Annual average power service outage duration per customer between 2013-2021 \cite{zhongming2021us}.]{\label{Data from 2020}{\includegraphics[width=0.45\textwidth]{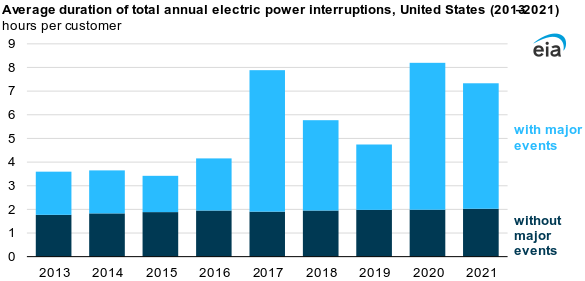}}}\hfill
\caption{Historical data on the frequency and the impact of power outages in the U.S.}
\end{figure}

The frequency and severity of power outages have increased as a result of several factors including a rise in extreme weather events, increasing demand for goods and services, and infrastructure network deterioration. \cite{hines2009large}, \cite{kenward2014blackout} and \cite{jufri2019state} present data related to large power failures due to extreme weather events between 1984 and 2006 in the U.S. In this period, 75\% of power outages were caused by earthquakes, ice-storms,  hurricanes, thunderstorms and tornadoes affecting more than 50,000 customers each time. Severe weather between 2003 and 2012 resulted in a recorded total of 679 outages in the U.S., impacting 10 million people.
Some of the major power outages due to weather-related events that occurred between 2016 and 2021 are shown in Table \ref{Tab:MajorOutages}. Researchers underscore the gravity of the situation by documenting 1.33 billion outage hours experienced by utility customers in 2020, representing a substantial 73\% increase over the previous year \citep{Hering}. Figure \ref{electrical dist in 2020} shows that in 2020, out of 400 electrical disturbances, 43\% were due to natural disasters. Over the years, electrical disturbances have consistently increased across all seasons. Supporting this trend, data from the U.S. Energy Information Administration, depicted in Figure \ref{Data from 2020}, corroborates an escalating average time that electrical power is unavailable or disrupted over a year due to natural disasters. In 2020, customers experienced an average power outage of approximately six hours due to major events. Prolonged power outages have severe economic consequences, affecting various sectors and activities. \cite{sanstad2020case} delve into the economic impact of power interruptions lasting from a minimum of a few days to an extended duration of several weeks due to extreme weather, drawing insights from six case studies. The escalating frequency of such events and their far-reaching consequences underscore the pressing need for proactive measures to enhance the resilience and reliability of power infrastructures.

\begin{table}[h]
\scalefont{0.8}
\centering
\caption{Major power outages worldwide between 2016 and 2021. }
\label{Tab:MajorOutages}
\begin{tabular}
{r| r |r | r  }

\hline \hline 
    \textbf{Date} & \textbf{Location}  & \textbf{Weather Event} & \textbf{Number of People Affected} \\ \hline
    September, 2016 & Australia & Tornado & 1,700,000 \\ \hline March, 2017 & U.S. & Winter Storm & 1,000,000 \\ \hline September, 2017 & Puerto Rico & Hurricane Maria & 300,000 \\
\hline October, 2017 & U.S and Canada &  Tropical Storm Philippe & 1,800,000 \\ \hline December, 2017 & U.S. &  Winter Storm Benji & 900,000 \\ \hline September, 2018 & Japan & Earthquake & 2,950,000 \\\hline December, 2018 & Canada & Windstorm & 600,000 \\ \hline June, 2019 & U.S. & Thunderstorm & 350,000 \\ \hline July, 2019 & U.S. & Tornado & 277,000 \\ \hline January, 2020 & Indonesia & Thunderstorm & 6,800,000 \\ \hline April, 2020 & U.S. & Tornado & 4,300,000 \\ \hline August, 2020 & U.S. & Tornado & 6,400,000 \\ \hline October, 2021 & Australia & Wind storm & 5,200,000 \\ \hline \hline
\end{tabular}
\end{table}
\normalsize

In the following sections, we review quantitative models that explore power network restoration strategies including\textit{ repair crew dispatching, pre-positioning emergency power sources, incremental network design, and network reconfiguration} measures. 

\subsection{Repair crew dispatching}
One course of action to restore a disrupted power network is dispatching utility crew(s) to the affected region to visit and repair faulty parts of the power network. The goal is to mobilize the crew to restore the network with minimum negative impact, which is a function of the power service outage duration, the affected population's size, and the type of interrupted activities (e.g., commercial or household). Service disruption time is determined by the crew's repair route sequence and the topology of the underlying network. Because of the inter-dependencies within a power grid, repairing a fault at a location does not ensure restoring service at the node unless all faults in predecessor nodes between the node and power source are repaired as well. In addition, when developing a repair schedule, changes in the electrical current within the network may need to be considered, especially in transmission systems, to avoid further damaging the network. Therefore, determining crew repair routes poses a considerably greater challenge compared to the classical 
Traveling Salesman Problem (TSP) and Vehicle Routing Problem (VRP) \citep{toth2014vehicle}. To handle this challenge, related studies address repair crew routing problems using different approaches. The first group uses a scheduling approach ignoring travel times while the second group models the problem as a routing problem that accounts for both service outage duration and crew travel times. We review the related studies within these two groups in the following subsections.

\subsubsection{Repair crew scheduling}
Studies that employ a scheduling approach for repair crew management problems commonly disregard travel times and assume identical repair crews, with one repair crew assigned to all damaged network components or multiple crews assigned to multiple damaged components.

Scheduling a single repair crew for restoring different power network structures can be traced back to \cite{guha1999efficient}. They minimize the weighted latency for 
different power network structures. 
In a tree power network, the authors formulate a dynamic program to find a connected sub-graph from a given graph $G$ such that the distance of all the vertices from the source in the sub-graph remains within a particular value. For bipartite power networks, they develop a logarithmic approximation method. They create a polynomial time algorithm and a greedy heuristic to find the solution for these tree and bipartite networks. \cite{tan2019scheduling} consider repair crew scheduling in power networks with a radial structure. The authors formulate a time-indexed integer linear programming (ILP) model with repair and network flow constraints, and compare the solution of the ILP with three proposed heuristic algorithms. The first algorithm is a linear programming (LP)-based list scheduling algorithm with soft precedence constraints. The soft precedence constraints consider the rank between the faulty nodes of the radial power network. The second algorithm converts the repair schedule of the single crew to multi-crew. Finally, the third algorithm measures the component's importance and dispatches the repair crew based on a set of rules. The second algorithm outperforms the other two and is computationally more efficient for real-time implementation. \cite{tan2020scheduling} extend the repair crew scheduling problem for radial power network structures to the case where the damage information is incomplete. The authors consider uncertainty in repair times. Their solution method uses an LP-based list scheduling policy, which requires statistical estimates of the repair times, such as mean and variance. The performance of the policy is flexible as it depends on knowledge of the distributions. \cite{xu2019improving} consider a multiple repair crew scheduling problem for recovering large-scale infrastructures with extensive disruptions under limited repair resources. They propose two novel heuristic methods combining time- and component-index-based methods to solve the problem. Their first heuristic reduces the optimality gap, while the second focuses on decreasing computation time. The proposed heuristics do not address the travel time between two repair tasks, inter-dependency between the different infrastructure networks, or budget constraints.

Considering the interdependency between a power network and another infrastructure network, the existing research on restoring interdependent infrastructure systems focuses on a sequential decision-making process (scheduling and repairing). However, there are significant drawbacks to using this approach because the restoration process involves multiple conflicting objectives, which means that optimizing one objective might negatively impact another, leading to sub-optimal solutions in terms of overall system restoration. Instead of performing sequential decisions, both \cite{coffrin2012last} and \cite{cavdaroglu2013integrating} propose a mixed-integer programming (MIP) model that integrates repair and the scheduling decisions. \cite{coffrin2012last} is the first to consider multiple repair crew scheduling for the final phase of the restoration process, which involves reconnecting the last segments of the networks to restore the entire functionality of interdependent power and gas networks. They modify the objective function of the VRP to develop an MIP model that jointly maximizes the load of the gas and power networks. The authors reduce the original MIP formulation to a smaller version by selecting important nodes and, subsequently, the restoration order of these nodes. They develop the randomized adaptive decomposition (RAD) algorithm that selects smaller restoration sub-problems and solves them independently to find effective and superior solutions within a limited time. \cite{cavdaroglu2013integrating} study a repair scheduling problem for interdependent electrical power and communication infrastructure systems. The authors develop an MIP-based heuristic that works in three phases. The first phase focuses on the operation of the highest priority nodes, the second phase selects the optimal set of arcs based on the interdependent layered network to make the power and communication systems function effectively together, and finally, the third phase is the scheduling phase, which schedules the crew sequence on these selected arcs. The heuristic approach provides high-quality scheduling decisions with computational resources that perform efficiently in a limited time. \cite{arab2015proactive}
propose a mixed-integer linear programming (MILP) model for assigning repair crews to damaged components for power network repair. The model in the response phase leverages the pre-disaster crew allocation solution as the initial solution and updates the repair crew schedule considering the impact of the disaster on the road network, which is predicted based on the intensity and the path of the disaster. 


\subsubsection{Repair crew routing}
In repair crew routing problems, single-crew and multi-crew problems are significantly different since the individual contributions of each repair route towards the objective function are not independent. \cite{salehipour2011efficient} investigates an efficient approach to restoring power networks following an outage. They frame the problem as a Traveling Repairman Problem to minimize the overall time during which services are disrupted. They develop a two-phase meta-heuristic algorithm to find a solution. The first phase is the construction phase, which uses a greedy randomized adaptive search procedure (GRASP), whereas the second phase is the improvement phase, which uses variable neighborhood descent. \cite{ccavdar2022repair} study a repair crew routing problem for a power distribution network. The authors develop an exact solution method by combining forward and backward dynamic programming approaches. Their method employs structural results that partially characterize suboptimal solutions. 

\cite{van2011vehicle} address power service recovery using multiple repair crew routing to minimize the total power outage time by separating routing optimization from power restoration. This decoupling generates two sub-problems of power restoration. The first sub-problem finds the minimum set of arcs to repair so that the system can be restored to full capacity, whereas the second sub-problem finds the order of restoration for these arcs, which results in a VRP, providing a joint repair schedule and route with high solution quality for large-size instances. \cite{morshedlou2021heuristic} consider routing multiple repair crews. The authors develop an MILP model and propose a new heuristic algorithm. The heuristic first uses a density-based mapping and a clustering algorithm, which clusters the disrupted locations based on their importance instead of the distance from the depot and accordingly allocates the repair crews to those locations. Then, it uses a local search procedure on the output of the first part to find a solution. They solve the problem for a power network with 289 components, a water network with 120 components, and a gas network with 33 components. \cite{morshedlou2018work} study the repair crew routing problem to maximize the power network's resilience, which they measure through a ``time-dependent function of recovered network performance compared to total performance loss." The authors develop two MIP models. The first model assumes binary restoration variables representing immediate service restoration, and the second model enhances the level of operation of links in a network that has undergone partial restoration. As the restoration process progresses, these links become more functional or operational. They also propose a relaxed version of MIP formulation, disregarding arrival times and crew effects on component restoration rates. The algorithm that solves this MIP model works in two steps. The first step eliminates different cycles in the graph to synchronize crew routes and resolve timing conflicts, while the second step calculates crew routing time and resolves conflicts caused by late arrivals. \cite{li2021hybrid} is the first study to consider a repair crew routing problem integrated with distribution network restoration under uncertainty (i.e., uncertainty in restoration time, component failures, distribution generator output and electrical demand) to reduce the total cost associated with both the operation of the system and any expenses incurred due to the loss of demand. 
This paper presents a multi-stage model. The first stage determines the repair crew's routing sequence before realizing the uncertainties. The second stage reconfigures the distribution network using stochastic programming, considering uncertainties in crew availability, component failures, and restoration time. Lastly, the third stage generates solutions robust to fluctuations in the power source output and demand conditions. The solution is obtained using column generation and progressive hedging approaches. 

\subsection{Dynamic location of portable energy sources }

In power infrastructure systems, distributed generators (DGs) and mobile or portable energy storage systems (MESS or PESS) play an important role in restoring service in a timely manner while the network is under repair. The amount of fuel reserve or battery charge within the network is usually limited after a disruption. 
When distributed generators are insufficient, mobile energy systems can be integrated into power grids. During this integration, network reconfiguration and maintaining the power balance in the network are critical. Integration decisions are done considering cost, which includes service interruption cost and network operation costs. \cite{yao1998repair} and \cite{yao2018transportable} discuss the feasibility of integrating MESS into a power grid in conjunction with repair crew location and dispatch decisions to minimize the outage duration and improve the efficiency of the power failure restoration processes. Routing and scheduling portable energy sources become increasingly complicated when these activities are performed in parallel with repair crew management. 
To handle this challenge, related studies use different approaches. The first group manages the dynamic location of the energy storage alone, whereas the second group integrates the positioning of the energy storage along with repair crew dispatching. In the following, we summarize related work within these two groups.

\emph{Dynamic location of portable energy sources alone: }\cite{wang2022novel} and  \cite{xu2019resilience} focus on routing for delivering portable energy systems. They identify the constraints representing the state transitions and  travel times for portable energy systems, which simplify to linear constraints under mild assumptions.  Using these constraints, the authors propose an MILP model to maximize the load restoration and minimize the travel cost. \cite{yao2019rolling} study scheduling and routing of mobile energy source systems after disrupting both power and road network networks. The authors consider a mobile energy system location problem to minimize the total cost, which combines customer service interruption and operational costs. To reduce the computation time, they formulate the scheduling of mobile energy systems using a probabilistic multi-tier spatiotemporal network.  To generate realizations of road status and load consumption random variables, they run a Monte Carlo simulation. The authors use rolling optimization to dynamically update the status of system damage for the generated realizations.
\cite{wang2022multi} formulate the routing and scheduling problem for portable energy storage systems as an incomplete information Markov game. Each MESS unit is treated as an autonomous agent in 
deep reinforcement learning algorithms to enable agents to learn optimal actions from the environment. They use a ``parameterized multi-agent double deep Q-learning network" that can handle various sources of uncertainty including energy storage capacity, traffic volume, demand-supply fluctuations, and network topology. \cite{coffrin2015transmission} consider co-optimizing load pickups and generator dispatching decisions to produce a repair sequence that minimizes the power outage over time. However, in most cases, generator dispatching uses DC power flow equations, which may cause violations of operational constraints.

Electric buses (EBs) are also used as temporary mobile energy storage systems. \cite{li2021routing} and \cite{su2022critical} develop dispatching plans for a cluster of EBs after a disaster. The EBs provide a unique advantage by not only serving as a means of transportation for affected customers but also supporting power demand during the restoration process. \cite{li2021routing} minimize the total downtime cost and energy consumption under different constraints related to network structure, electricity usage and current distribution. They formulate an MILP model without considering uncertainties in the weather and the status of the road, and use CPLEX to solve the optimization problem, significantly reducing outage costs. \cite{su2022critical} study the use of EBs for power distribution system restoration. They first determine the EB routes based on the network's traffic and subsequently develop a load restoration model. The authors then integrate the uncertainties in the number and locations of faults in the power network, along with road congestion affecting the road network, into these models as a constraint to formulate an EB routing model. They convert the uncertainty constraints into deterministic form and use a ``second-order conic relaxation" and ``piecewise linearization" to obtain the solution. \cite{lei2016mobile} and \cite{lei2018routing} propose a two-stage dispatch framework for distributed generators to minimize the expected downtime considering node priorities and demand sizes of the loads. The first stage involves pre-positioning resources, e.g., allocating repair crews and repair equipment, including the distributive generators, whereas the second stage is real-time allocation. \cite{lei2016mobile} use a shortest path algorithm and pre-disaster information to allocate generators in real-time after a disruption. \cite{lei2018routing} formulate the problem as a mixed-integer second-order cone programming model and transform it into an MILP.

\emph{Dynamic location of energy sources and repair crew dispatching: } \cite{arif2017power} present a two-stage method for dynamic positioning of a mobile energy source and repair crew dispatching. In the first stage, the authors propose an integer program to cluster damaged components based on the repair crew's availability and distances of the faulty nodes from the depot. In the second stage, they integrate MESS dispatching and repair crew routing to minimize the time required for repairs and maximize the efficiency of load collection. The authors treat repair crew routing as a VRP with constraints on resources and repair crew attributes. \cite{lei2019resilient} formulate a non-convex mixed-integer non-linear programming (MINLP) model to solve three subproblems: distribution system restoration, repair crew dispatch and mobile energy source dispatch. The distribution system restoration depends on the availability of mobile energy sources and repair crews for dispatching. The model is non-convex and converted to a mixed-integer second-order cone program and then linearized to create an MILP. The simultaneous positioning of mobile energy source systems and repair crew dispatching for radial power networks is extended by \cite{arif2018optimizing} to a case where uncertainties exist in node demands and repair times. They develop a two-stage stochastic MILP where the repair crews visit the faulty nodes in the first stage and generators reconfigure the power network in the second stage. The authors decompose the problem into two stochastic sub-problems to maximize the load and dispatch the repair crews to the damaged components. A progressive hedging algorithm solves these two subproblems unless all faulty nodes are repaired. \cite{ding2020multiperiod} consider the coordination of mobile battery-carried vehicles, repair crews, and network microgrids for power restoration. The authors formulate the problem as an MILP and solve it using CPLEX. They introduce an acceleration algorithm to optimize the MILP model by dynamically adjusting the objective function and improving computational speed while preserving solution accuracy within the original feasible region obtained using CPLEX. 

\subsection{Incremental network design and network reconfiguration}

In this section, we review response strategies focusing on the network topology to restore a disrupted power network. Incremental network design and network reconfiguration fall within this group.

\subsubsection{Incremental network design}
\textit{Incremental network design} involves adding new arcs to a network to potentially use in the response phase to increase network capacity under a limited budget. This strategy allows for the potential utilization of the newly added arcs to enhance the network's capabilities. However, implementing incremental network design is a complex task, as modifying the arc structure in power networks can overload the network's non-faulty sections, potentially leading to larger power outages. Researchers in the field have addressed incremental network design using various approaches, focusing on load balancing after the addition of new arcs to the network to avoid such issues and ensure a smooth power flow.

\cite{kalinowski2015incremental} study incremental network design problems to optimize the aggregate flow within a finite planning horizon. They present two MIP formulations and three heuristic solution methods. The first heuristic efficiently augments the flow and is a $\frac{3}{2}$-approximation algorithm for a network with capacity constraints. The second heuristic is a 2-approximation algorithm that helps the network connections reach the optimized flow faster. The third heuristic combines both strategies, allowing for network adjustments when necessary and lowering the risk of being stuck with a rigid network setup. 
\cite{binato2001new} solve incremental power network design and expansion problems using a variation of Benders' decomposition. The method uses a linear (0–1) disjunctive model, derives the minimum values for disjunctive parameters to address numerical difficulties, and computes the coefficients of the Benders' cuts. The advantage of decomposition is that it helps attain convergence. Gomory cuts are integrated within the Benders' decomposition framework to further improve convergence. \cite{baxter2014incremental} use a shortest path objective to address incremental network design problems. To solve smaller networks, they use an integer programming model, whereas, for more extensive networks, they develop a greedy heuristic with a worst-case performance guarantee, and propose a 4-approximation algorithm. 

\subsubsection{Network reconfiguration}

\textit{Network reconfiguration} allows using on/off switches on power lines to change the network topology. By doing so, some nodes can be served immediately. This approach proves beneficial in reducing the overall impact of power outages. For instance, \cite{chen2017modernizing} propose a decision support tool that incorporates different restoration plans for network reconfiguration, including routing distributed generators, switching, and employing a simulator for micro-grid formation. However, network reconfiguration poses its challenges, as it requires careful consideration to avoid overloading the non-faulty areas and ensure a reliable power supply. Studies in the current literature address network reconfiguration using three different approaches. The first group performs network reconfiguration using mobile power sources and network partitioning. The second group conducts network reconfiguration using distribution generators. The third group integrates network reconfiguration with repair crew dispatching decisions. We review the related studies below.

\emph{Network
reconfiguration using mobile power sources, switching operations and network partitioning: } \cite{huang2017integration} propose a two-stage robust MILP for network reconfiguration using network switching mechanisms and dispatching portable energy storage systems for service restoration after a disaster. The authors use a nested column-and-constraint-generation decomposition technique to find a solution. In particular, switching mechanisms prove beneficial in maintaining power balance and reducing load shedding during emergencies. \cite{wang2016resilience} discuss the presence of uncertainty in component failures. They formulate a recursive Markov model for network reconfiguration considering the impact of topology changes in the network. The different topologies form the state space in the model, and the component failure rates constitute the transition probabilities. The probability of each state determines the weight of each objective function individually, which helps to transform the model into an MILP model that is solved using CPLEX. \cite{samudrala2020distributed} propose a novel outage detection algorithm for more decentralized control in large distribution networks where scalability and communication are significant issues. The authors follow a network partitioning approach for network reconfiguration and calculate a solution for each sub-network in parallel. \cite{bienstock2007using} address the problem of preventing the cascading effect of a large-scale power outage using network reconfiguration. The authors propose two models. The first model formulates an MIP that determines node capacity investments that ensure demand satisfaction at all nodes under any failure scenario.  The authors use a branch-and-cut algorithm with a Benders' cut generation scheme to solve the MIP exactly, while employing rounding and standalone heuristics to obtain good feasible solutions.  The second model considers the detailed dynamics of successive rounds of cascading power failures (and the associated lost edges in the network).  A given set of capacities is considered \emph{survivable} if it meets some minimum proportion of total demand under any scenario. An LP subproblem determines power flows that minimize the deviation from the power delivered in previous round, which determines whether a proposed set of capacities is survivable. The authors use a cutting plane algorithm to determine optimal capacity investments that ensure survivability. \cite{liu2020survivability} study the stable operation of smart grids to detect power outages and repair a disrupted power network by reconfiguring the topology while informing operators of faulty locations. They propose a deep reinforcement learning framework with a reward function for calculating the routing path. The authors use an improved deep Q network for faster convergence. \cite{thiebaux2013planning} consider a sequence of switching operations to reconfigure a faulty network where the primary objective is to maximize the number of customers receiving power service, while at the same time minimizing the number of network switching operations. 
They propose a mixed-integer mathematical formulation that provides a high quality switching sequence for an arbitrary topology within a given time constraint. 

\emph{Network reconfiguration using distribution generators: }
\cite{chen2015resilient}, \cite{wang2015research}, and \cite{xu2016microgrids} study network reconfiguration on power distribution systems to reduce downtime and enhance operational continuity help prevent delays, maintain product quality, and ensure timely deliveries, which are critical for customer satisfaction and operational efficiency. They form multiple micro-grids energized by distribution generators (DGs). \cite{chen2015resilient} formulate the micro-grid formation problem as an MILP to maximize the total loads that are considered to be high priority while satisfying each micro-grid connectivity constraint, branch node constraint, load pickup constraint, operation constraint and topology constraint. They design a multi-agent coordination scheme to collect operational parameters and network topology information through local wireless communication between switches and facilitate decentralized decision-making for forming multiple microgrids using a consensus algorithm. 
\cite{xu2016microgrids} perform dynamic simulations assuming a limit on the amount of energy generated by the micro-grids. \cite{chen2017sequential} formulate an MILP model with the help of DGs and switching operations that maximizes the total restored energy over a time horizon considering the weights of the loads and constraints, including DG operational constraints, connectivity, topological, and sequencing constraints.

\emph{Network reconfiguration with repair crew dispatching: }\cite{ding2020multiperiod} propose an MILP model that co-optimizes the dispatching of repair crews, mobile energy systems, and network micro-grids for power restoration, where network micro-grids are used for network reconfiguration. The authors minimize the total weighted loads while incorporating constraints on the travel and repair times of the repair crew, balancing the demand and supply of power-grids, and load demands. \cite{arab2016electric} develop an optimal repair schedule and network reconfiguration plan for restoring a disrupted power network grid. The value of the lost load measures the importance of each load in the power grid. The authors formulate an MIP model and use a Benders' decomposition algorithm to find the solution. \cite{chen2018toward} formulate an MILP that considers the routing sequence of repair crews, network reconfiguration through remote-controlled and manual switches, and load energizing sequence. \cite{carvalho2007dynamic} propose a set of network reconfiguration operations performed by a repair crew. The optimal sequence of these operations is determined using a dynamic programming approach. Since the number of states increases exponentially, the authors employ two methods for state reduction. In the first method, assessment of the state transitions involving significant travel time is ignored. In the second method, the evaluation of state transitions is confined to an arbitrary order. \cite{van2015transmission} maximize the customer count receiving power after a disruption through a two-stage model. The first stage develops an MIP model for the power flow equations, which is non-linear and non-convex. The second stage uses extensive neighborhood search and randomized adaptive decomposition to find the repair crew route. The output of the first stage generates precedence relationships, which act as an input for the second stage. \cite{nurre2012restoring} formulate an IP model along with valid inequalities to maximize the cumulative weighted power flow arriving at the demand nodes on a power network. The authors develop a heuristic algorithm that creates a set of rules or guidelines for the repair crew decisions on which network arcs to restore  at each step of the restoration process.

\section{Road Network Restoration}
\label{Sec:Road}
Transportation networks facilitate critical operations including the movement of mobile medical facilities, food, temporary shelter, and search and rescue operations after a disruption. After a disaster, road networks may become disconnected as a result of factors including debris, fallen trees and buildings, displaced cars, and structural damages to roads, which hinder relief operations. The level of disruption increases as the road networks age and become more vulnerable. Developing efficient restoration strategies can become especially challenging due to uncertainties (e.g., the amount of demand and supply for different items at various locations, damage to roads limiting access to certain areas, travel times), and limitations (e.g., delivery time restrictions). These challenges have motivated several studies. We classify studies on road network restoration based on their strategic emphasis area, namely repair crew dispatching, incremental road network design, and debris clearance. In the following sections, we review related literature within these categories.

\subsection{Repair crew dispatching}

One course of action to restore a disrupted road network is by dispatching utility crew(s) to the affected region to repair faulty edges of the road network. The goal is to restore connectivity between locations and facilitate the movement of food, medical care and people as quickly as possible. Studies address repair crew dispatching problems using
different approaches. The first group generally ignores the travel distances and models the problem as a repair crew scheduling problem. 
The second group models the problem as a repair crew routing problem. 

\subsubsection{Repair crew scheduling}

We distinguish between single- and multi-crew problems, since the contributions of individual repair crew schedules with respect to the objective function are not independent in the multi-crew case, and generalizing solution approaches effective for single-crew problems is not straightforward.

\textit{Single-crew scheduling problems for road network restoration} have similarities to the Traveling Repairman Problem (TRP). TRP solution methods focus on minimizing latency or travel time by determining the repair crew route, and can be applied within the context of road network restoration. These methods can help prioritize which sections of the road network should be restored first and then determine how to allocate resources efficiently. \cite{silva2012simple} develop a meta-heuristic algorithm for the minimum latency problem, which uses a combination of three algorithms: variable neighborhood descent, which explores different neighborhoods around the current solution to avoid getting stuck in a local optima, local search, and GRASP. The algorithm can quickly generate optimal or near-optimal solutions on small instances in a fraction of a second and improves the solutions for larger instances. \cite{bulhoes2018branch} propose a branch-and-cut algorithm that aims to identify partitions of a set.  The subsets created during the partitioning process are disjoint, ensuring that each feasible solution is distinct and does not overlap, satisfying the constraints and objectives of the minimum latency problems. \cite{akbari2021weighted} study a modified version of the online minimum latency problem incorporating restrictions on using specific connections between nodes in the network and including weights in the nodes, suggesting that each location has a particular significance or cost. They develop two polynomial-time heuristics, namely, an ``Adapted Greedy" and a ``Shortcut Algorithm," to address and solve the problem.  \cite{duque2016network} study the problem of scheduling emergency repair crews for rural road networks damaged by natural disasters to reach locations requiring medical and food assistance. This study presents two distinct approaches to solving the problem. For small to medium network sizes, the authors use dynamic programming, whereas when the network instance is medium- to large-scale, they use iterative GRASP. The study underscores the significance of strategic and well-thought-out approaches to repair and recovery activities, offering managerial guidance for more effective and informed decision-making during the post-disaster period.

For \textit{multi-crew repair dispatching problems}, two fundamental aspects that significantly impact decision-making are the nature of repair and travel times, and the composition of the repair crews. We categorize and analyze existing research based on these two critical dimensions:

\emph{Nature of travel and repair times}: The nature of repair and travel times can vary widely, ranging from deterministic to stochastic. In the deterministic setting, \cite{yan2007time} is the first to study time-space network flow models for road restoration by constructing an integer programming model. The central goal of the study is to reduce the overall waiting time for emergency repair operations. To solve larger instances, they decompose the the problem into sub-problems and use a heuristic algorithm and CPLEX to for their solution. In a separate study, \cite{yan2012ant} address the flow of entities across a road network and the temporal aspects associated with their movement over time. They introduce an ant-colony-system algorithm employing stochastic search techniques to find the solution. Notably, while efficient, this model needs to account for the stochasticity of the time needed for travel and repairing the nodes. Furthermore, if there is a need to introduce more repair teams to address the situation, the existing plan needs to adequately consider the impact on the original work schedule. Building on the concept of minimizing the total waiting time for emergency repair operations, \cite{yan2014optimal} consider disruptions in the availability of goods, services and resources due to a series of disasters in a short time. They formulate an MILP model with three distinctive features: the flow of entities across a road network and the temporal aspects associated with their movement over time for dynamic network updates, variations from the original schedule, and repair team formation during each variation period. The authors employ an ant-colony-system-based hybrid global search algorithm to find the solution. However, it does not explicitly address stochastic repair and travel times. \cite{tang2009short} study short-term teamwork scheduling under uncertain repair and travel times. Their model introduces penalty and semi-deviation risk measures for these stochastic scenarios, formulating two objective functions; one reduces the total time to travel and fix the nodes, including penalties and semi-deviation risk measures, while the other reduces the expected values of these measures. 

\emph{Composition of repair crew}: Repair teams can be either homogeneous or heterogeneous. Homogeneous crews consist of members with similar skills and capabilities, allowing for interchangeable assignments. In contrast, heterogeneous crews involve members with diverse skills or specialized roles, requiring strategic allocation based on task requirements. \cite{moreno2019branch} present an MIP for dispatching multiple homogeneous crews and develop a branch-and-cut algorithm to find a solution. The scheduling decision is treated as the master problem and the routing decision as the corresponding subproblem. The subproblem is solved using a shortest path algorithm. \cite{moreno2020heterogeneous} consider multiple heterogeneous crews. To model the schedule and routes of the heterogenous repair crews, the authors formulate three MIP models. The first two of these are 2-index and 3-index VRPs with additional valid inequalities, while the third formulation eliminates some variables from the 2-index VRP formulation and adds new constraints to prevent the restoration of damaged nodes that do not affect the accessibility of demand nodes, which improves the computation performance compared to the original 2-index formulation.
Comparing the solution of all three models, the 2-index VRP model performs better than the other two. The models do not consider uncertainty in the travel or repair times. \cite{moreno2020decomposition} develop an enhanced branch-and-cut algorithm as an exact solution method for repair crew scheduling. The authors improve the computational efficiency by introducing innovative ways of organizing decision variables and imposing additional constraints in their formulation. Their solution approach uses a genetic algorithm and a simulated annealing algorithm. Finally, the authors combine the Benders' cut approach with meta-heuristics to develop a hybrid solution that further improves the solution the meta-heuristic solution. The models do not consider uncertainty in the status of the damaged nodes. \cite{iloglu2020maximal} study the inter-dependency between the repair crew and emergency service responders aiming to maximize emergency demand coverage by coordinating road repairs and emergency responder relocations over time. The nodes represent demand points and facility locations, while arcs represent paths between them. Recovery crews repair damaged arcs, and emergency responders are relocated using these arcs to maximize coverage. They develop an integer programming formulation and two heuristics. The first heuristic uses Lagrangian and linear programming relaxations. The second heuristic applies integer rounding to the solution obtained by the linear programming relaxation.

\subsubsection{Repair crew routing}

Repair crew routing not only determines the sequence for equipment pickup and repair activities, but also affects the amount of information obtained about the true state of a disrupted road network in incomplete information situations. 
We classify the related papers into two main groups. The first group considers repair crew routing based on the VRP. The second group approaches repair crew routing based on the cumulative capacitated VRP. Another stream of literature formulates these problems using dynamic programming, although this literature is limited. One notable work using dynamic programming as the solution approach is \cite{ulusan2021approximate}. The authors study the repair crew routing problem under uncertainty, and  formulate this problem as a Markov decision process, using approximate dynamic programming for its solution. They approximate the value functions as a linear combinations of states, where each state is considered as an independent variable, and the coefficients represent the weights assigned to these states. The linear combination of state values, each multiplied by a weight, forms the estimated value function. This approach enables them to derive near-optimal policies. Their model does not consider crew inter-dependencies, demand uncertainty, node locations or debris amounts.

\textit{Vehicle Routing Problem (VRP) Approach for Repair Crew Routing: }\cite{kim2006waste}, \cite{buhrkal2012waste}, and \cite{zhang2017manpower} study road restoration for facilitating the transfer of medical supplies and food using a VRP approach. 
\cite{kim2006waste} develop a mathematical programming model considering route compactness, which typically involves shorter distances and fewer turns that further reduce the travel time. The authors use a construction algorithm to determine the routes. To address the computational challenges in large problem instances, they propose a clustering algorithm. Similarly, \cite{buhrkal2012waste} formulate a mathematical programming model to minimize the crew travel cost under various constraints and employ an adaptive large neighborhood search algorithm to find a solution. \cite{zhang2017manpower} contribute by introducing an integer programming model and utilizing a variable neighborhood search strategy to find a solution. \cite{akbari2017multi} study the logistical challenge of efficiently routing repair crews in a coordinated way to restore a road network, with a specific emphasis on minimizing the time needed to clear disrupted edges. They formulate an exact MIP and relaxed MIP models to solve small and large instances. A local search-based heuristic algorithm solves the relaxed MIP. \cite{akbari2017multii} study the dispatching of repair crews to maximize the creation of access points to isolated communities or groups of people within the road network after disruptions. Similarly to \cite{akbari2017multi}, they formulate exact and relaxed MIPs, and use a Lagrangian method to solve the relaxed MIP. The authors also develop two mathematical programming-based heuristic algorithms to obtain a solution and a lower bound. 
\cite{akbari2021decomposition} study how to route multiple crews in the aftermath of a disruption. The authors formulate an MIP model and use local search heuristics to obtain a solution. Before using the algorithm, they compute initial solutions, which are crucial in reducing the computation time. \cite{kasaei2016arc} study repair crew routing under two scenarios. In the first one, certain arcs in the network are assigned a higher priority, while in the second one, all arcs are treated in the same way throughout the network restoration process. The authors propose two algorithms. The first is a variable neighborhood search algorithm, and the second is a variable neighborhood search algorithm combined with construction heuristics.


\cite{barbarosoglu2004two}, \cite{mete2010stochastic}, \cite{caunhye2016location} and \cite{goldbeck2020optimal} study road restoration for medical supply distribution after a disaster. The models and approaches they have adapted for road network restoration scenarios address different situations after a disaster, such as identifying strategic locations for storing road construction materials and equipment or determining the quantity and types of materials needed at different locations. \cite{barbarosoglu2004two} consider vehicle routing formulations for road restoration in two stages. In the first stage (preparedness stage), they determine each depot's location and inventory level. In the second stage (response stage), a stochastic programming model minimizes the total costs and transportation time. The model proposed by \cite{mete2010stochastic} minimizes the preparedness cost in the first stage, and the second stage minimizes the total restoration duration and penalty for any unmet restoration objectives or incomplete tasks. \cite{caunhye2016location} minimize the preparedness cost in the first stage, while the second stage minimizes the total response time. \cite{goldbeck2020optimal} present a multi-stage stochastic programming model to improve supply chain resilience. The model tries to minimize the total supply chain cost, which also includes penalties for unmet demand for improved resilience, by deciding the investments and operational planning in the pre-disruption phase, and recovery activities in the post-disruption. It integrates capacity planning, operational adjustments, and recovery strategies within a unified framework, which allows for the dynamic allocation of repair resources. A case study demonstrates the model's ability to quantify the benefits of pooling repair resources, thereby improving the overall resilience of supply chains. Unlike traditional sequential models \cite{zhang2022dynamic} develop an integer programming model that performs inspection and restoration simultaneously. They dynamically update the repair crew routes based on the information revealed over time. The authors propose a hybrid genetic algorithm solution approach that combines a traditional genetic algorithm with a specific way of representing solutions and a heuristic method that stops the process early under certain conditions to improve computational efficiency. 

\textit{Cumulative Capacitated Vehicle Routing Problem (CCVRP) Approach for Repair Crew Routing: } In general, the CCVRP differs from the classical VRP by finding a set of vehicle routes that minimizes the total customer arrival times while ensuring
the capacity constraints are met and each customer or location within a cluster is visited exactly once. In road network restoration problems, the cumulative service disruption time considers the problem from the affected population's perspective. Therefore, several researchers study the repair crew routing problem using this approach. \cite{ngueveu2010effective} study repair crew routing for a homogeneous fleet to reduce the total arrival times at the faulty nodes during relief operations after a disaster. Their solution methodology combines a genetic algorithm with local search to obtain reasonable lower bounds. \cite{ribeiro2012adaptive} study a VRP in the aftermath of a disruption to reduce the total arrival times at each node. The authors develop a search algorithm, which employs removal and reinsertion heuristics and an adaptive search engine that adjusts these heuristics. The engine uses a probabilistic selection method, where past performance determines the likelihood of choosing a heuristic. Additionally, the authors employ a strategy to adjust the weights of these heuristics based on recent performance in consecutive steps. The algorithm performs better as compared to the genetic algorithm. \cite{lysgaard2014branch} study multiple repair crew routing to reduce the cost related to the average arrival time over the set of damaged nodes for a certain route. The authors formulate a CCVRP consisting of subsets of all potential routes and 2-index vehicle flow formulations. They simplify this complex problem using linear programming relaxation, transforming it into a Dantzig-Wolfe master problem and finally using a branch-and-cut algorithm for the solution. Similarly, \cite{luo2014branch} consider the problem under a maximum travel distance constraint. Their solution uses a different variant of the branch-and-cut algorithm in which the forward and backward search algorithm optimizes the column generation pricing sub-problem. \cite{rivera2016mathematical} study repair crew routing for road restoration when a single vehicle makes multiple trips according to a solution to the CCVRP. They develop two MILP models. One model is based on the flow of resources on each of the arcs to minimize the total arrival time at  the disrupted nodes, and the other involves dividing the problem into subsets and using commercial solvers for solving instances up to 20 nodes. The authors also reformulate the multiple-trip CCVRP as a resource-constrained shortest path problem and use an algorithm for finding solutions in a weighted graph containing up to 40 nodes. \cite{wang2019effective} consider a multi-depot CCVRP for distributing disaster relief materials and fresh food. They use a local search algorithm to find the solution. \cite{cinar20162} provide a modified variant of the  \cite{clarke1964scheduling} algorithm to minimize the distance travelled and the load carried by the repair vehicle. They also propose a $k$-means clustering algorithm along with a two-phase construction heuristic to improve the computational performance of the CCVRP.

\subsection{Incremental road network design}
Incremental network design is also a common strategy for road network restoration. Incremental network design for road networks involves (i) adding new arcs or (ii) increasing existing arc capacities to ensure network connectivity for faster deliveries to affected locations. The former strategy is called \emph{discrete} and the latter is called \emph{continuous} network design \citep{meng2002benefit}. Studies addressing incremental network design problems mainly focus on heuristic approaches, which can be classified into two groups. The first group models the problem assuming complete information about the damaged arcs and demands, whereas the second group models the problem assuming incomplete information. We next review the related studies in these two groups. 

\emph{Complete Information: } \cite{meng2002benefit} study the continuous network design problem after a disruption to create a balanced and impartial recovery where the distribution of resources and benefits is fair and inclusive for all communities. They formulate a bi-level programming model and use simulated annealing to find the solution. \cite{murawski2009improving} study a discrete incremental network design problem in a healthcare setting to maximize the number of residents with access to critical facilities. Instead of essential healthcare facilities, the focus could be on strategically placing resources such as construction equipment, materials, and workforce to maximize the efficiency of road restoration. They formulate an MILP model and use CPLEX for its solution. \cite{scaparra2005grasp} consider a network design problem focusing on arc capacity decisions to improve traffic flow in the aftermath of a disruption. Their objective function maximizes connectivity and minimizes the total weighted travel distance. The authors combine GRASP with a heuristic algorithm to connect different paths to improve the solution. They further enhance the solution by using reactive GRASP, allowing for better parameter calibration and re-linking the existing solution. \cite{karamlou2014optimal} formulate a multi-objective combinatorial optimization problem for restoring bridges to reduce the travel time for connecting important locations and maximize resilience, which they measure based on the percentage of bridges in service. For the solution, the authors use genetic algorithms, which provide a better restoration sequence and convergence compared to the methodology in \cite{bocchini2013computational}. \cite{zhao2020transportation} propose a bi-objective optimization framework for restoring a road network to minimize travel times and unmet demand. The upper-level problem studies  management actions to restore the edges of the disrupted road, whereas the lower-level problem studies the actions of the users who travel on the road network. The solution of the upper-level problem iteratively identifies and updates the set of active constraints. Similarly, the solution of the lower-level problem iteratively finds the minimizer of a linear approximation of the objective function.  

\emph{Incomplete Information: } \cite{chen2004stochastic} consider a stochastic setting for the network design problem and develop two stochastic models. The first model considers uncertain demand; the second model creates a balanced and impartial recovery focusing on equitable distribution of resources. The authors generate uncertain demands using simulation and use a genetic algorithm to find a solution. \cite{dimitriou2008reliable} study the network design problem as a stochastic bi-level programming problem, aiming to minimize the travel and expenditure costs in the upper-level problem under monetary constraints, network reliability, and physical constraints. The lower level problem details how changes in the network's capabilities, determined at the upper level, influence how users respond and adapt their demands. The authors employ an iterative approach using Monte Carlo simulation to estimate the parameters. As the solution approach, they use a genetic algorithm. \cite{ukkusuri2009multi} study the network design problem focusing on demand stochasticity and uncertainty in transportation networks. They formulate a single-stage mathematical program from a traditional two-stage network design formulation to maximize consumer surplus. They use sequential quadratic programming to solve the problem. In addition, the authors evaluate the single-stage mathematical program formulation by computing the value of flexibility and a flexibility index for this flexible network design problem. \cite{chen2010stochastic} formulate a multi-objective bi-level programming model for road capacity expansion decisions under demand uncertainty. The lower-level sub-problem determines the rules that users follow when there is an increase in network capacity combined with uncertain or unpredictable demand for road network. Meanwhile, the upper-level component involves a stochastic multi-objective model with three variations designed to capture the entire road network demand uncertainty under budget constraints. The authors combine genetic algorithms, multi-objective optimization, and simulation methods to find solutions to this complex problem with multiple objective functions.

\subsection{Debris clearance}

Debris makes parts of the road network inaccessible after a disaster, obstructing logistics operations for critical services and products. Therefore, clearing debris in a timely manner becomes a crucial task for transportation network restoration. Debris clearing includes several activities, such as hauling, collecting, transporting, disposal and recycling. 
Various studies address debris clearance problems using two different approaches. The first group models the problem assuming complete information is available on the amount and location of the debris, and the second group models the problem assuming incomplete information. We next review related studies in these two groups.

\emph{Complete Information: } \cite{ozdamar2014coordinating} study the assignment and routing of vehicles for cleanup operations after a disruption to minimize the cleanup time and maximize a network's overall reach and connectivity. They formulate an MIP model and use construction heuristics to help in finding efficient solutions by considering different factors and rules, such as priorities and cluster of blocked edges, during the solution-building process. They gauge the reliability and suitability of the heuristics under specific and known cleanup time conditions. \cite{sahin2016debris} consider a model that separately minimizes the time it takes to remove the debris and the total travel time required for the repair crew to visit all critical nodes. The authors use a shortest path algorithm to find an initial route and apply a variation of the 2-opt algorithm to improve the solution. \cite{ajam2019minimizing} study vehicle routing for the debris removal problem and develop a model to minimize the total waiting time experienced at critical nodes. The authors develop a metaheuristic approach that uses GRASP in the construction step and an iterative variable neighborhood search to improve the solution. \cite{berktacs2016solution} develop two MIP models and use construction heuristics to solve debris clearance problems. The first model minimizes the total time required to visit  critical nodes by applying a shortest path algorithm. In the second model, each path receives a weight or priority. The model aims to minimize the weighted sum of the visit times at critical nodes. Therefore, the second model provides a more nuanced approach, considering both the efficiency of the route and the significance of each critical node. To improve the quality of the solutions obtained from the construction heuristics, they use a 2-opt algorithm. \cite{briskorn2020simultaneous} study how to simultaneously clear roads and deliver relief goods to satisfy demand after a disruption. The authors leverage the structural features of the problem by transforming it into a more manageable form with a reduced graph. They then formulate an MIP model based on this reduced representation and employ a branch-and-cut algorithm to precisely solve the problem, taking advantage of its inherent structure for computational efficiency. \cite{yan2009optimal} formulate a bi-objective MIP model to minimize the debris removal duration and the time it takes to deliver relief goods to the demand points. The authors adopt a weighted objective, assigning a weight to each criterion based on importance and combining these into a single objective function. They use a decomposition strategy to break down the extensive problem into more manageable subproblems, which are then efficiently solved using CPLEX. Additionally, they introduce a construction heuristic to determine initial solutions. \cite{markov2016integrating} consider debris removal after a disruption and formulate an MILP by integrating a heterogeneous fleet and flexible depot assignment into the VRP, using valid inequalities to enhance solution efficiency. Their solution approach proposes a multiple neighborhood search heuristic by exchanging the positions of two elements in the solution, removing an element from its current position, reinserting it at a different location in the solution, and swapping two edges to improve the solution. \cite{liu2021optimization} study debris clearance with a bi-objective model that minimizes operational cost (e.g., transportation, maintenance, fuel costs) and environmental cost (e.g., release of carbon compounds or greenhouse gases from the burning or decomposition of debris). They formulate the problem as a VRP and use $k$-means clustering with Clarke \& Wright's \citeyearpar{clarke1964scheduling} algorithm to find an initial solution, which is improved using a search method.

\emph{Incomplete Information: } In addition to considering debris clearance under complete information,  \cite{ozdamar2014coordinating} also evaluate the effectiveness of construction heuristics under stochastic cleanup times. They perform a simulation study to measure the performance of their algorithm in detecting the blocked edges when the cleanup times are uncertain. \cite{ccelik2015post} and \cite{fikar2018agent} study the stochastic debris clearance problem to maximize the total expected relief supply post disruption. \cite{ccelik2015post} formulate the incomplete information problem as a partially observable Markov decision process where certain aspects of the system are not directly observable, introducing uncertainty into decision-making. To solve the problem, they use a tree structure where each node represents a state of the problem, and branches emanating from nodes represent possible actions. The authors eliminate certain branches or possibilities from the search tree based on heuristic criteria to enhance computational efficiency, reducing the overall search space. \cite{fikar2018agent} develop a simulation-based solution framework. They create an agent-based simulation to minimize cost and maximize services delivered (i.e., distribution of relief goods) under stochastic conditions. \cite{akbarpour2021innovative} and \cite{elgarej2017distributed} study  debris clearance under uncertainty. The authors formulate a VRP and leverage meta-heuristic approaches. Specifically, \cite{akbarpour2021innovative} employ simulated annealing, while \cite{elgarej2017distributed} explore ant colony optimization and a genetic algorithm to derive effective solutions for the debris clearance problem. \cite{sayarshad2020dynamic} consider a system with incomplete information to position equipment for debris clearance after a disruption. They formulate the problem as an MDP and propose an approximate dynamic programming approach.   \cite{yacsa2022metaheuristics} formulate a two-stage multi-period stochastic model for debris clearance, assuming zero lead time for resource allocation in the aftermath of a disruption. They establish an efficient and effective sequence for repairing the broken links in the initial stage. The next stage involves optimizing the allocation of repair crews to specific links using the established repair sequence from the initial stage. The authors propose two meta-heuristics. The first is a variable neighborhood search method, whereas the second combines tabu search and simulated annealing. 

\section{Recovery of Water, Gas and Oil Networks}
\label{Sec:Others}
In addition to power and road networks, water, gas and oil networks are also critically important for residential and industrial activities. Studies focusing on the logistics of the recovery of water, gas and oil networks are sparse compared to power and road networks. Therefore, we review related papers for each of these network types within this section.

\textit{Water network restoration:} Water plays an important role in maintaining an ecological balance within modern society. However, water networks suffer from natural disruptions that lead to huge losses and disrupt residential, commercial, and industrial activities.

\cite{todini2000looped} is the first to study the resiliency of water networks after a disruption. They formulate a multi-objective optimization problem to minimize the repair cost and maximize the water network's resilience, which they measure by the energy used at the water network's source. The author generates the Pareto set representing the optimal trade-offs between minimizing repair costs and maximizing resilience. 
Building on this study, \cite{creaco2016combined} aim to enhance the reliability of the water network at a low cost by introducing new connections corresponding to additional links or pathways established within the water network. This expansion of the multi-objective optimization problem aligns with network design considerations, emphasizing the importance of optimizing network structure for improving resiliency and efficiency post-disruption. In essence, the work of \cite{creaco2016combined} contributes to the broader field of network design strategies, particularly in the context of water infrastructure. Similarly, \cite{piratla2013performance} focus on simulating states involving single and double pipe breaks to assess water network performance. They use resilience metrics that are based on the inherent capacity or robustness of the water network to withstand and recover from failures or increased demand. The authors formulate a multi-objective model to minimize repair costs and maximize resilience. Their solution methodology uses a genetic algorithm that reconfigures the network by adjusting the layout, connections, or operational parameters to optimize the performance of the water network under new circumstances. \cite{farahmandfar2017resilience} study the recovery of a water network using network reconfiguration as a course of action. The authors introduce a resilience metric based on the pipeline capacity and network topology and develop an optimization framework subject to budget constraints to improve resilience at minimum cost. They use a binary genetic algorithm for its solution. \cite{tabucchi2010simulation} study repair of water networks after a disaster. They use a simulation model to schedule a repair crew for inspection and simultaneously execute repairs for the damaged network post-disaster. The authors compare their actions during the restoration process with the decisions made during an actual event showing reasonable accuracy. However, the model development is disruption-specific, and  finding a solution is computationally demanding. 
\cite{liu2020recovery} develop a demand-based resilience metric that considers how well the water network can meet or adapt to varying levels of demand during and after a disruptive event. Their solution methodology uses a genetic algorithm to obtain a high-quality pipe recovery sequence. \cite{morshedlou2018work} study repair crew routing to maximize the long-term robustness and sustainability of the water network. They measure resilience through a ``time-dependent function of recovered network performance compared to total performance loss." The authors formulate two mixed integer programming models. One of the models assumes binary restoration representing immediate service restoration, and the other assumes proportional restoration representing gradual service restoration. \cite{nozhati2020optimal} study the stochastic recovery of water networks to minimize the number of days required for restoration and maximize the number of people with access to utility services. The authors formulate the problem as an MDP and use approximate dynamic programming techniques, namely a rollout algorithm, to find the solution.

\textit{Gas and oil networks restoration:} 
\cite{he2017gas} study gas network recovery and formulate a non-linear integer multi-objective model, which minimizes the total cost, including recovery time cost, unmet demand cost, gas generation cost, and transmission cost.  
Their solutions reveal the balance between the cost of recovery time and the loss associated with unmet demand.
\cite{atsiz2022coordinated} study a single repair crew routing problem to restore interdependent gas and power networks after a disruption. They formulate an MIP model to reduce the total recovery time of all damaged components, and use a constructive heuristic and a simulated annealing algorithm to find a solution.  \cite{ahmed2020resilient} focus on establishing a resilient oil network system post-disruption using concepts from network design. The authors propose a strategy for service management through strategic actions like placing, replicating, and migrating data and services. They employ a game-theoretic approach with a two-player, zero-sum game, where player one represents the pipeline monitoring system, and player two represent the element failures. The goal is to improve system availability, resilience, and scalability, considering factors such as energy consumption, latency, and node storage capacity.

\section{Summary and Future Work Directions}

\label{Sec:Conc}
This paper reviews literature on quantitative studies in infrastructure network restoration, focusing primarily on the associated logistics operations. Our review considers network restoration for (i) power, (ii) road, and (iii) water, oil and gas infrastructures. For each group, we classify related studies based on the most common disruption response strategies. For power network restoration, we review studies on repair crew dispatching, dynamic location of portable energy source systems, incremental network design, and network reconfiguration. For road network restoration, we review repair crew dispatching, incremental network design and debris clearance. Since relatively fewer papers discuss recovery strategies for water, gas and oil infrastructure networks, we review these categories based on infrastructure network type. We emphasize the computational challenges of timely and immediate decisions as events unfold during the disruption response phase and highlight the difficulty of simultaneously achieving both speed and high-quality solutions. The review methodology involves a thorough search of peer-reviewed studies over the past 20 years. Tables \ref{comparison table1}, \ref{comparison table2} and \ref{comparison table3} in Appendix \ref{App:Summary} provide a comprehensive overview of the reviewed papers, presenting the key aspects considered in the related literature.

The literature on power network restoration emphasizes the increasing frequency and severity of power outages in recent years. Several studies highlight the impact of power service disruptions, including the duration and number of affected customers. Repair crew dispatching, a crucial decision for reconnecting disconnected power networks, is studied with respect to single- and multi-crew scheduling problems. Various models address repair crew assignment problems, considering factors such as road network conditions and the intensity and paths of disasters. Researchers use several methodologies, including stochastic programming and robust optimization, to minimize system operating costs and demand curtailment costs in stochastic repair crew routing. Another restoration strategy involves integrating distributed generators and portable energy storage systems into power grids for efficient service restoration. Addressing more complex issues, researchers study the coordination of routing and scheduling of these portable energy sources in conjunction with repair crew dispatching. 
Network topology modification is another approach for restoring disrupted power networks, focusing on incremental network design and reconfiguration. We summarize the literature on network reconfiguration in conjunction with  portable energy source location decisions, network partitioning, deployment of distribution generators, and repair crew dispatching.

In road network restoration, the most common decisions involve repair crew dispatching, incremental road network design, and debris clearance. Repair crew dispatching, a crucial decision for reconnecting disconnected road networks, is characterized in terms of single- and multi-crew scheduling problems. Notably, the nature of the repair and travel times and crew heterogeneity add complexity to repair crew routing problems. Incremental road network design is another strategy that is studied under complete and incomplete information. The third group of road network restoration studies focuses on debris clearance decisions under different objectives, such as minimizing the cleanup time and total travel time needed to clear roads and deliver relief goods. 
An additional group of studies proposes two-stage multi-period stochastic models that integrate resource allocation, repair sequence determination, and crew scheduling under uncertainty. 

The recovery of water, gas, and oil networks after disruptions is crucial for maintaining domestic and industrial activities. While studies in this area are fewer than those considering power and road networks, the existing research addresses various aspects. In water network restoration, multi-objective optimization models
focus on minimizing repair costs and maximizing network resilience. Gas network recovery studies involve nonlinear integer programming models that balance recovery time costs and unmet demand costs. Coordinated repair crew routing strategies explore interdependent gas and power networks. 

Based on the review of the current state-of-the-art methods on infrastructure network restoration, one important research direction lies in developing models that account for both incomplete information on network status and the precedence relationships between different tasks or repairs. 
Such an integrated approach will provide a more realistic representation of real-world scenarios. 
An important area for future research is the exploration of interdependencies between various infrastructure networks. 
Investigating how different networks, such as power and road networks, influence each other during repair processes can lead to more efficient and effective solutions. 
Existing literature primarily concentrates on power and road networks. Exploring how these studies extend to water, gas and oil networks, as they differ from general networks in terms of functionality, infrastructure characteristics, environmental impact, and interdependencies, would be valuable.
By addressing these multiple dimensions simultaneously, future research can provide a holistic and realistic framework for infrastructure network restoration in a complex and dynamic environment.

\section{A Brief Comparison of the Reviewed Papers}

\label{App:Summary}
In this section, Tables \ref{comparison table1}, \ref{comparison table2} and \ref{comparison table3} provide a summary of some of the reviewed papers based on the characteristics of the underlying problem studied.

\begin{table}[H]

\centering
\caption{Summary of the reviewed papers for power network restoration.}
\label{comparison table1}

\resizebox{\textwidth}{!}{\begin{tabular}
{ c|c|c|c|c|c} 
\hline \hline

\textbf{Paper} &\textbf{Crew Dispatching}  &\textbf{Incremental Network Design} 
 & \textbf{Mobile Source Positioning} & \textbf{Stochastic} & \textbf{ Precedence}\\
\hline 
\cite{arab2015proactive}
& $\bullet$ &   &   &  & $\bullet$ \\\hline 
\cite{arab2016electric}
& $\bullet$ & $\bullet$   &   & & $\bullet$  \\\hline \cite{arif2017power}
& $\bullet$ &   &   &  & $\bullet$\\\hline \cite{arif2018optimizing}
& $\bullet$ &   &   & $\bullet$  & $\bullet$\\\hline 
\cite{baxter2014incremental}
&  & $\bullet$ &  &   & $\bullet$ \\\hline
\cite{bienstock2007using}
&  & $\bullet$ &  &   & \\\hline 
\cite{binato2001new}
&  & $\bullet$ &  &   & \\\hline        
\cite{carvalho2007dynamic}
& $\bullet$ & $\bullet$ & & $\bullet$    &\\\hline
\cite{ccavdar2022repair}
& $\bullet$ &  &  &  & $\bullet$\\\hline
\cite{cavdaroglu2013integrating}
& $\bullet$ &  &  &  & \\\hline
\cite{chen2018toward}
& $\bullet$ &  $\bullet$ &  &    &  $\bullet$\\ \hline \cite{chen2017sequential}  
&  & $\bullet$ &  &    & $\bullet$ \\\hline \cite{chen2015resilient}
&  & $\bullet$ &  &    & $\bullet$ \\\hline 
\cite{chen2017modernizing}
&  & $\bullet$ &  &  & $\bullet$  \\\hline  
\cite{coffrin2012last}  
& $\bullet$ &  &  &  & \\\hline \cite{coffrin2015transmission}
&  & $\bullet$ &  &   & $\bullet$ \\\hline   
\cite{ding2020multiperiod}
&  &  & $\bullet$ &  & $\bullet$\\\hline 
\cite{guha1999efficient} 
& $\bullet$ &  & &   & \\\hline
\cite{huang2017integration}
&  & $\bullet$ & &   & $\bullet$\\\hline  
\cite{kalinowski2015incremental}
&  & $\bullet$ &  &    & $\bullet$\\\hline  \cite{lei2016mobile}
&  & $\bullet$ &  &   & $\bullet$ \\\hline  \cite{lei2018routing}
&  & $\bullet$ &  &  & $\bullet$ \\\hline  \cite{lei2019resilient}
& $\bullet$ & $\bullet$ &  &  & $\bullet$ \\\hline 
\cite{li2021routing}
&  &  & $\bullet$ &  &$\bullet$ \\\hline  \cite{liu2020survivability}
& &  & $\bullet$ &    & $\bullet$\\\hline
\cite{li2021hybrid}
& $\bullet$ &  &   & $\bullet$  & $\bullet$ \\\hline
\cite{morshedlou2018work}
& $\bullet$ &  &  &   & $\bullet$\\\hline  
\cite{morshedlou2021heuristic}
& $\bullet$ &  &  &   & \\\hline
\cite{nozhati2020optimal}
& &  & $\bullet$ &  $\bullet$  & \\\hline
\cite{nurre2012restoring}
& $\bullet$ & $\bullet$  &  &    & $\bullet$\\\hline
\cite{salehipour2011efficient}  
& $\bullet$ &  & &  & $\bullet$  \\\hline \cite{samudrala2020distributed}
& & $\bullet$ &  &  & $\bullet$\\\hline 
\cite{su2022critical}
&  &  & $\bullet$ &  & $\bullet$\\\hline 
 \cite{tan2020scheduling}
& $\bullet$ &  &  & $\bullet$  & \\\hline 
\cite{thiebaux2013planning}
&  & $\bullet$ &  & &  $\bullet$ \\\hline 
\cite{van2011vehicle}
& $\bullet$ &  &  &   & \\\hline  
\cite{van2015transmission}
& $\bullet$ &  &  &   & $\bullet$ \\\hline  
\cite{wang2015research}
&  & $\bullet$ & &    &$\bullet$ \\\hline \cite{wang2016resilience}
&  & $\bullet$ & &    &$\bullet$ \\\hline
\cite{wang2022novel}
&  &  & $\bullet$ & & $\bullet$\\\hline 
 \cite{wang2022multi}
&  &  & $\bullet$ & $\bullet$  & \\\hline   
\cite{xu2016microgrids}
&  &$\bullet$  &  &  & \\\hline 
\cite{xu2019improving}
& $\bullet$ &  &  &   & \\\hline 
 \cite{xu2019resilience}
&  &  & $\bullet$ &  & $\bullet$\\\hline 
\cite{yao1998repair}
&  &  & $\bullet$ &  & $\bullet$\\\hline 
 \cite{yao2018transportable}
&  &  & $\bullet$ &  & $\bullet$\\\hline
\cite{yao2019rolling}
&  &  & $\bullet$ &  & $\bullet$\\\hline  \hline
\end{tabular}}
\end{table}

\begin{table}[H]

\centering
\caption{Summary of the reviewed papers for road network restoration.}
\label{comparison table2}

\resizebox{\textwidth}{!}{\begin{tabular}
{ c|c|c|c|c|c} 
\hline \hline
\textbf{Paper} &\textbf{Crew Dispatching} & \textbf{Incremental Network Design} & \textbf{Debris Collection} & \textbf{Stochastic}  & \textbf{Precedence}\\
\hline 
\cite{ajam2019minimizing} &  &  & $\bullet$ &  & $\bullet$\\ \hline 
\cite{akbari2021decomposition}  
& $\bullet$ &  & &  &   \\
\hline
\cite{akbari2017multi}
& $\bullet$ &  & &  & $\bullet$\\\hline 
\cite{akbari2017multii}
  & $\bullet$ &  & &  & $\bullet$\\\hline
  \cite{akbari2021weighted}
& $\bullet$ &  &  &  & \\\hline  
\cite{akbarpour2021innovative}
&  &  & $\bullet$ & $\bullet$  & $\bullet$\\\hline
\cite{barbarosoglu2004two}
& $\bullet$ &  & &  &    \\\hline     \cite{berktacs2016solution}
& &  & $\bullet$ &  & $\bullet$  \\\hline
\cite{bocchini2013computational}
& &$\bullet$   & &  & $\bullet$\\\hline 
\cite{briskorn2020simultaneous}
&  &  & $\bullet$ &  &  $\bullet$  \\\hline 
\cite{bulhoes2018branch}
& $\bullet$ &  & &  &  \\\hline  
\cite{buhrkal2012waste}  
& $\bullet$ &  &  &  & \\\hline 
\cite{caunhye2016location}
& $\bullet$ &  & &  &  \\\hline
\cite{ccelik2015post}
&  &  & $\bullet$ &  $\bullet$  & $\bullet$ \\\hline
\cite{chen2004stochastic}
&  &$\bullet$  &  &  $\bullet$  & $\bullet$\\\hline
\cite{chen2010stochastic}
& & $\bullet$ &  & $\bullet$  & $\bullet$\\\hline
\cite{cinar20162}
& $\bullet$ &  & &  &  \\\hline
\cite{dimitriou2008reliable}
&  & $\bullet$ &  & $\bullet$  & $\bullet$\\\hline
\cite{duque2016network}  
& $\bullet$ &  &  &  &  \\\hline '
\cite{elgarej2017distributed}
&  &  & $\bullet$ &  $\bullet$  & $\bullet$\\\hline 
\cite{fikar2018agent}
& &$\bullet$   &  &  &  $\bullet$\\\hline 
\cite{goldbeck2020optimal}
& $\bullet$ &  & &  & \\\hline
\cite{han2015waste}
& $\bullet$ &  & &  &  \\\hline
\cite{iloglu2020maximal}
& $\bullet$ &  &  &  & \\\hline 
\cite{karamlou2014optimal}
& & $\bullet$   & &  &  $\bullet$  \\\hline 
\cite{kasaei2016arc}
& $\bullet$ &  & &  & $\bullet$ \\\hline 
\cite{kim2006waste}
& $\bullet$ &  & &  &  \\\hline  
\cite{liu2021optimization}  
&  &  & $\bullet$ &  &  $\bullet$\\\hline
\cite{luo2014branch}
& $\bullet$ &  &  &  &   \\\hline \cite{lysgaard2014branch}
& $\bullet$ &  &  &  &\\\hline  
\cite{markov2016integrating}
&  &  & $\bullet$ &  & $\bullet$ \\\hline
\cite{meng2002benefit}
& & $\bullet$ &  &  & $\bullet$ \\\hline 
\cite{mete2010stochastic}
& $\bullet$ &  & &  $\bullet$  & \\\hline \cite{moreno2019branch}
& $\bullet$ &  & &  & $\bullet$\\\hline \cite{moreno2020heterogeneous}  
& $\bullet$ &  & &  & $\bullet$ \\\hline \cite{moreno2020decomposition}
& $\bullet$ &  &  &  & $\bullet$\\\hline 
\cite{murawski2009improving}
& &  $\bullet$ &  &  & $\bullet$\\\hline
\cite{ngueveu2010effective}
& $\bullet$ &  & &  &  \\\hline 
\cite{ozdamar2014coordinating}
&  &  & $\bullet$ &  & $\bullet$  \\\hline
 \cite{ribeiro2012adaptive}
& $\bullet$ &  & &  &  \\\hline \cite{rivera2016mathematical}  
& $\bullet$ &  & &  & \\\hline 
\cite{sahin2016debris}
&  &  & $\bullet$ &  & $\bullet$ \\\hline 
\cite{sayarshad2020dynamic}  
&  &  & $\bullet$ &  $\bullet$  & $\bullet$\\\hline 
  \cite{scaparra2005grasp}
& &  $\bullet$ &  &  & $\bullet$\\\hline 
\cite{silva2012simple}
& $\bullet$ &  & &  &  \\\hline 
\cite{tan2019scheduling}
& $\bullet$ &  & &  &  \\\hline 
\cite{tang2009short}
& $\bullet$ &  &  &  &\\\hline   
 \cite{ukkusuri2009multi}
& &  $\bullet$ &  & $\bullet$  & $\bullet$\\\hline 
\cite{ulusan2021approximate}
& $\bullet$ &  &   & $\bullet$  & \\\hline
\cite{wang2019effective}
& $\bullet$ &  & &  & \\\hline

\cite{yan2007time}
& $\bullet$ &  & &  & \\\hline
\cite{yan2009optimal}
&  &  & $\bullet$ &  & $\bullet$\\\hline
\cite{yan2012ant}
& $\bullet$ &  & &  &  \\\hline 
\cite{yan2014optimal}
& $\bullet$ &  & &  &\\\hline
\cite{yacsa2022metaheuristics}
&  &  & $\bullet$ & $\bullet$  & $\bullet$\\\hline 
 \cite{zhang2017manpower}
& $\bullet$ &  & &  & \\\hline
 \cite{zhang2022dynamic}
& $\bullet$ &  &  &  &  \\\hline
\cite{zhao2020transportation}
&  & $\bullet$ & &  & $\bullet$\\ \hline \hline 
  
\end{tabular}}
\end{table}

\begin{table}[H]

\centering
\caption{Summary of the reviewed papers for water, gas and oil network restoration.}
\label{comparison table3}
\resizebox{\textwidth}{!}{\begin{tabular}
{c|c|c|c|c} 

\hline \hline
\textbf{Paper} &\textbf{Crew Dispatching} & \textbf{Network Design} & \textbf{Stochastic}  & \textbf{Precedence}\\
\hline 
\cite{ahmed2020resilient}
&  & $\bullet$ &     & \\\hline 
\cite{atsiz2022coordinated}
&  & $\bullet$    &     & \\\hline 
\cite{creaco2016combined}
&  & $\bullet$  &    & \\\hline
\cite{farahmandfar2017resilience}
& &  $\bullet$  &      & \\\hline 
\cite{he2017gas}
&  & $\bullet$  &     & \\\hline 
\cite{liu2020recovery}
&  & $\bullet$  &    & \\\hline 
\cite{morshedlou2018work}
& $\bullet$ &    & & $\bullet$\\\hline
\cite{morshedlou2021heuristic}
& $\bullet$ &   &    & \\\hline
\cite{nozhati2020optimal}
  &  & $\bullet$   &$\bullet$  \\\hline
\cite{piratla2013performance}
& & $\bullet$ &       & \\\hline 
\cite{tabucchi2010simulation}
& & $\bullet$  &     & \\\hline 
\cite{todini2000looped}
&  & $\bullet$  &    & \\\hline \hline
\end{tabular}}
\end{table}

\normalsize

\bibliographystyle{apalike}

\bibliography{review.bib}

\appendix

\end{document}